\documentclass[11pt,letterpaper]{scrartcl}

\usepackage[margin=1in]{geometry}
\usepackage{microtype}
\usepackage{etoolbox}
\usepackage{subcaption}
\usepackage{booktabs}

\usepackage{tikz}
\tikzstyle{fsquare} = [rectangle,inner sep=2pt,outer sep=3pt,fill]
\tikzstyle{fcircle} = [circle,inner sep=1.5pt,outer sep=3pt,fill]
\tikzstyle{hcircle} = [circle,inner sep=1.5pt,outer sep=3pt,draw]
\usetikzlibrary{decorations.pathreplacing,calligraphy,math}

\usepackage[ruled,vlined]{algorithm2e}

\usepackage{aligned-overset}
\usepackage{color}

\definecolor{darkgreen}{rgb}{0.0 0.66 0.0}

\title{Tighter  Approximation for the  Uniform Cost-Distance Steiner Tree Problem}

\usepackage{amssymb,amsmath,amsthm, amsfonts}
\usepackage{hyperref}

\newtheorem{theorem}{Theorem}[section]
\newtheorem{lemma}[theorem]{Lemma}
\newtheorem{definition}[theorem]{Definition}

\author{Fine Foos, Stephan Held, and Yannik Kyle Dustin Spitzley\\
  Research Institute for Discrete Mathematics and\\
  Hausdorff Institute for Discrete Mathematics\\
  University of Bonn\\
  \{foos,held,spitzley\}@dm.uni-bonn.de}

\begin{document}

\maketitle

\begin{abstract}
  Uniform cost-distance Steiner trees minimize the sum of the total length and
  weighted path lengths from a dedicated root to the other terminals.  They are applied
  when the tree is intended for signal transmission, e.g.\ in chip
  design or telecommunication networks. They are a special case of
  general cost-distance Steiner trees, where different distance functions are
  used for total length and path lengths.

  We improve the best published approximation factor for the uniform
  cost-distance Steiner tree problem from $2.39$ \cite{Khazraei21} to
  $2.05$. If we can approximate the minimum-length
  Steiner tree problem arbitrarily well, our algorithm achieves an
  approximation factor arbitrarily close to $1+\frac{1}{\sqrt{2}}$.
  This analysis is tight.
  We also prove the gap
  $1+\frac{1}{\sqrt{2}}$ between optimum solutions and the lower bound which
  we and all previous approximation algorithms for this problem use.

  Similarly to previous approaches, we start with an approximate minimum-length
  Steiner tree and split it into subtrees that are later reconnected.
  To improve the approximation factor, we split it into components more carefully,
  taking the cost structure into account, and we significantly enhance the analysis.
   We also show that no algorithm using a pure split and reconnect strategy can achieve an approximation factor better than $\frac{3}{2}$.
\end{abstract}

\section{Introduction}

Steiner trees can be found in numerous applications, in particular in
chip  design and telecommunications.  In these applications,
both the total tree length and the signal speed are important.  We
consider Steiner trees that do not only minimize the total cost, but also the weighted path lengths from a dedicated root $r$ to the
other terminals.
Formally, the  problem is defined as follows.

An instance $(M,c,T,r,p,w)$ consists of a metric space $(M,c)$, a root $ r$, a finite set $ T $ of sinks, a map $p:T\dot{\cup}\{r\}\to M$,  and sink delay weights $ w\colon T\to \mathbb{R} _{\geq 0}$.
The task is to compute a  Steiner tree $ A $ for $ T\cup \{ r\} $ with an extension $p: V(A) \setminus (T\cup\{r\}) \to M$ minimizing
\begin{align}
  \label{eq:objective}
  \sum _{\{x,y\}\in E(A)} c(p(x),p(y)) + \sum _{t\in T} \left(w(t)\!\!\!\! \sum_{\{x,y\} \in E(A_{[r,t]})} \!\!\!\!c(p(x),p(y))\right),
\end{align}
where $ A_{[r,t]} $ is the unique $ r $-$ t $-path in $ A $. We call (\ref{eq:objective}) the \textbf{(total) cost} of $(A,p)$.

Given a Steiner tree $A$, we call
$$\sum _{\{x,y\}\in E(A)} c(p(x),p(y)) \quad\text{ its \textbf{connection cost} and }$$
$$\sum _{t\in T} \left(w(t)\sum_{\{x,y\} \in E(A_{[r,t]})} c(p(x),p(y))\right) \quad\text{ its \textbf{delay cost}. }$$

Usually the position $p$ of vertices  is clear from the
context. Then, we simply write $c(x,y)$ instead of $c(p(x),p(y))$ and
$c(e)$ instead of $c(x,y)$ for edges $e=\{x,y\}$.
To shorten the notation, we often also omit the underlying metric space from the notation
and write only $(T,r,w)$ to denote an instance.
A simple lower bound for the objective function, is given by
\begin{align}
  \label{eq:lower_bound}
  C_{SMT}(T\cup\{r\}) + D(T,r,w),
\end{align}
where $ C_{SMT}(T\cup\{r\})$ is the connection cost  of a minimum-length Steiner tree for $T\cup\{r\}$, i.e.\ a Steiner tree $ A $ for $ T\cup \{ r\} $  minimizing $ \sum _{e\in E(A)} c(e) $, and $D(T,r,w):=\sum_{t\in T} w(t) c(r,t)$
is the sum of weighted root-sink distances.

The \textsc{Uniform Cost-Distance Steiner Tree Problem}  was first mentioned by \cite{meyerson2008cost}, who considered the
(general) cost-distance Steiner tree problem, where the connection cost
may be unrelated to the delay cost.  Cost-distance Steiner trees are heavily  used in
VLSI routing and interconnect optimization \cite{held2017global,daboul2019global}.
Here, the weights arise as Lagrangean multipliers when optimizing global signal delay constraints on an integrated circuit \cite{held2017global}.
Uniform cost-distance Steiner trees are computed as a first step of a Steiner tree oracle in global routing \cite{held2017global,daboul2019global}.

The general
cost-distance Steiner tree problem does not permit an approximation factor better
than $\Omega(\log \log |T|)$ unless $\text{NP} \subseteq
\text{DTIME}(|T|^{\mathcal{O}(\log\log\log |T|)})$ \cite{Chuzhoy08}, while a randomized
$\mathcal{O}(\log |T|$)-factor approximation algorithm was given by \cite{meyerson2008cost} and \cite{Chekuri01}.
Meyerson, Munagala and Plotkin \cite{meyerson2008cost} observed  that a constant factor approximation
algorithm  for the \textsc{Uniform Cost-Distance Steiner Tree Problem}  can be obtained using the  shallow-light spanning tree algorithm from \cite{khuller1995balancing}.
The resulting factor is $3.57$. Using shallow-light Steiner trees \cite{held2013shallow} instead of spanning trees, the factor was improved
to $2.87$ independently by \cite{guo2014approximating} and \cite{rotter2017timing}.
The first algorithm using the delay weights  algorithmically  was given by Khazraei and Held \cite{Khazraei21}.
They achieve  an approximation factor of $ 1+\beta $, where $ \beta $ is the approximation factor for computing a minimum-length Steiner tree.
All these approaches compare against the lower bound in (\ref{eq:lower_bound}).

Similarly to algorithms for shallow-light trees, the algorithm in \cite{Khazraei21} starts from an approximately minimum Steiner tree,
which is cut into a forest whose components are connected to the root $r$ individually.
While \cite{khuller1995balancing}  cut the tree  whenever the path length is too large,
\cite{Khazraei21} cut off a subtree if its delay weight exceeds a certain threshold.
Each cut-off tree is later reconnected through a direct connection from the root
through one of its terminals, minimizing the resulting objective function.

The special case where we require a spanning tree instead of a Steiner tree and $w(t)$ is identical for all $t\in T$ is known as the
\textit{cable-trench problem}. It does not permit a PTAS unless $\mathrm{P}=\mathrm{NP}$ \cite{benedito2021inapproximability}.

The \textsc{Uniform Cost-Distance Steiner Tree Problem} is related to the \textit{single-sink buy-at-bulk problem}
where a set of demands needs to be routed from a set of sources to a single sink using a pipe network that has to be constructed from
a finite set of possible pipe types with different costs and capacities \cite{guha2009constant,talwar2002single,jothi2009improved}.
The best known approximation factor for this problem is  $40.82$ due to \cite{grandoni2010network}, who
also achieve a factor of $20.41$ for the splittable case.
If there is only one pipe type this problem is equivalent to the  \textsc{Uniform Cost-Distance Steiner Tree Problem}.
In fact, the threshold-based tree cutting used in the proof of \cite{Khazraei21}  is similar to the algorithm in  \cite{guha2009constant},
but the reconnect to the root/sink differs.

\subsection{Our contribution}

In this paper, we improve the approximation algorithm in  \cite{Khazraei21} for the
\textsc{Uniform Cost-Distance Steiner Tree Problem}.

\begin{theorem}
  \label{thm:main}
  The \textsc{Uniform Cost-Distance Steiner Tree Problem} can be approximated in polynomial time with an approximation factor of
\[\beta + \frac{\beta}{\sqrt{\beta ^2 +1} + \beta - 1},\]
  where $\beta \ge 1$  is the  approximation guarantee for the minimum-length Steiner tree problem.
\end{theorem}

With the best known approximation factor for the minimum Steiner tree problem $\beta = \ln(4)+\epsilon$ \cite{Byrka13,traub2022local}, this results in an  approximation factor  $< 2.05$
and for  $\beta = 1$ this  gives the factor $1 + \frac{1}{\sqrt{2}} <  1.71$, clearly improving upon the previously best  factors
$2.39$ and $2.0$ in  \cite{Khazraei21}.
The polynomial-time approximation scheme by \cite{Arora98} allows choosing $\beta$ arbitrarily close to one in the Euclidean and the Manhattan planes. However, general metric spaces do not allow $\beta \le \frac{96}{95}$ unless $\mathrm{P}=\mathrm{NP}$ \cite{chlebik2008steiner}.

Assuming an ideal Steiner tree approximation factor of $\beta=1$, our new approximation factor is tight
with respect to the lower bound  (\ref{eq:lower_bound}). We prove the following result
\begin{theorem}
  \label{thm:lb_gap}
  \[\sup_{T,c,w}\frac{\mathrm{OPT}(T,r,w)}{C_{SMT}(T\cup\{r\})  + D(T,r,w)}= 1 + \frac{1}{\sqrt{2}},\]
\end{theorem}
where $\mathrm{OPT}(T,r,w)$ denotes the optimum solution value for $(T,r,w)$.

The algorithm in  \cite{Khazraei21} starts from a short Steiner tree and iteratively splits off  subtrees whose  delay weight exceeds a given threshold.
We proceed similarly, but we also take the structure of the subtrees into account and split off subtrees once they can be reconnected efficiently.

While \cite{Khazraei21} obtain a running time of $\mathcal{O}(\Lambda + |T|^2)$,
where $\Lambda$ is the time to compute an initial $\beta$-approximate minimum Steiner tree,
our running time is  $\mathcal{O}(\Lambda + |T|)$ (assuming that the metric $c$ can be evaluated in constant time).
Thus, it is very fast and  applicable for large chip design instances.

We would like to mention that in a preliminary unpublished paper we
achieved worse factors of $2.15$ ($\beta=\ln(4)$) and $1.80$ ($\beta = 1$)
\cite{spitzleyheld22} using a more complicated algorithm and
analysis.
That paper also shows  that the factor $1+\beta$ in \cite{Khazraei21} is tight
for that algorithm.

The analysis of our algorithm is tight. Furthermore, no cut-and-reconnect
algorithm that starts with a minimum-length Steiner tree and cuts it into pieces
which are reconnected to the root $r$ can guarantee an approximation
ratio better than $\frac{3}{2}$.

The remainder of this paper is structured as follows.
In Section~\ref{sec:optimality_gap}, we show that the supremum in Theorem~\ref{thm:lb_gap} is at least $1 + \frac{1}{\sqrt{2}}$.
Then, in Section~\ref{sec:approximation-algorithm} we will briefly summarize the algorithm and proof from   \cite{Khazraei21}, as our work enhances it.

Our  improved splitting algorithm and analysis is presented  in Section~\ref{sec:improvement}.
The proof of Theorem~\ref{thm:main} is presented in Section~\ref{sec:main-result}.
It also shows that the supremum in Theorem~\ref{thm:lb_gap} is at most $1 + \frac{1}{\sqrt{2}}$.
The tightness of the analysis is shown in Section~\ref{sec:tightness_of_analysis}.
In Section~\ref{sec:worst-case-cut-reconnect}, we show the lower bound on the approximation factor
of any cut-and-reconnect algorithm.
We finish with conclusions in Section~\ref{sec:conclusion}.

\section{Optimality Gap of Lower Bound}
\label{sec:optimality_gap}
In this section, we will show that the gap between an optimum solution
and the lower bound in (\ref{eq:lower_bound}) can be as large as $1 +
\frac{1}{\sqrt{2}}$. Together with the  approximation factor of our
new algorithm for $\beta = 1$ (Theorem~\ref{thm:main}), the gap is asymptotically $1 + \frac{1}{\sqrt{2}}$.

\begin{theorem}\label{thm:lowerbounddeviation}
    There are instances $I^{(k)}_{k\in\mathbb{N}}$ with $I^{(k)} = (T^{(k)},r^{(k)},w^{(k)})$ $(k\in\mathbb{N})$ for the uniform  cost-distance Steiner tree problem such that
   \begin{equation*}
      \lim_{k\to\infty}\frac{\mathrm{OPT}^{(k)}} {C^{(k)} + D^{(k)}} = 1 + \frac{1}{\sqrt{2}} \text{,}
   \end{equation*}
   where $\mathrm{OPT}^{(k)}$ is the optimum value for the instance $I^{(k)}$,  while $C^{(k)}= C_{SMT}(T^{(k)}\cup \{r^{(k)}\}) $,  $D^{(k)}:=D(T^{(k)},r^{(k)},w^{(k)})$
   denote minimum possible connection cost and the minimum possible delay cost of $I^{(k)}$.
\end{theorem}
\begin{proof}
   \begin{figure}
      \centering
          \begin{subfigure}{0.6\textwidth}
              \centering
              \begin{tikzpicture}[scale=0.9]
                \node[fsquare,red,label=$r$] (r) at (0,0) {};

                \node[fcircle,label=right:\footnotesize{\textcolor{blue}{$\frac{1}{\sqrt{2}}$}},label=left:\footnotesize{$t_{k}$}] (1) at (320:4) {};
                \node[fcircle,label=right:\footnotesize{\textcolor{blue}{$\frac{1}{\sqrt{2}}$}},label=left:\footnotesize{$t_{k-1}$\!\!\!}] (2) at (290:3) {};
                \node[fcircle,label=left:\footnotesize{\textcolor{blue}{$\frac{1}{\sqrt{2}}$}},label=right:\footnotesize{$t_{2}$}] (3) at (250:3) {};
                \node[fcircle,label=left:\footnotesize{\textcolor{blue}{$\frac{1}{\sqrt{2}}$}},label=right:\footnotesize{$t_{1}$}] (4) at (220:4) {};

                \node[fcircle,label=below:$c$] (c) at (0,-5) {};

                \draw[-] (r) -- (1) node[midway,xshift=6pt,yshift=4pt]{$1$};
                \draw[-] (r) -- (2) node[midway,xshift=6pt,yshift=4pt]{$1$};
                \draw[-] (r) -- (3) node[midway,xshift=-6pt,yshift=4pt]{$1$};
                \draw[-] (r) -- (4) node[midway,xshift=-6pt,yshift=4pt]{$1$};

                \draw[dashed,-] (c) -- (r) node[midway,right] {\footnotesize{$2$}};
                \draw[dashed,-] (c) -- (1) node[midway,below right] {\footnotesize{$\frac{1}{\sqrt{2}}$}};
                \draw[dashed,-] (c) -- (2) node[midway,xshift=-6pt,yshift=6pt] {\footnotesize{$\frac{1}{\sqrt{2}}$}};
                \draw[dashed,-] (c) -- (3) node[midway,xshift=6pt,yshift=6pt] {\footnotesize{$\frac{1}{\sqrt{2}}$}};
                \draw[dashed,-] (c) -- (4) node[midway,below left] {\footnotesize{$\frac{1}{\sqrt{2}}$}};
              \end{tikzpicture}
              \caption{Illustration of the complete instance.}
              \label{fig:lowerbounddeviation}
          \end{subfigure}
          \hfill
          \begin{subfigure}{0.35\textwidth}
            \centering
            \begin{tikzpicture}[scale=0.75]
              \node[fsquare,red,label=$r$] (r) at (0,0) {};
                  \node[fcircle] (v1) at (0,-1) {};
                  \node[fcircle] (v2) at (0,-2) {};
                  \node (v3) at (0,-3) {$ \vdots $};
                  \node[fcircle] (v4) at (0,-4) {};
                  \node[fcircle] (w1) at (-3.4641/5,-5+3/5) {};
                  \node[fcircle] (w2) at (-2*3.4641/5,-5+2*3/5) {};
                  \node[rotate=49.107] (w3) at (-3*3.4641/5,-5+3*3/5) {$ \vdots $};
                  \node[fcircle] (w4) at (-4*3.4641/5,-5+4*3/5) {};

                  \node[fcircle,label=left:\footnotesize{\textcolor{blue}{$\frac{1}{\sqrt{2}}$}},label=right:\footnotesize{$t_i$}] (4) at (210:4) {};

                          \node[fcircle,label=below:$c$] (c) at (0,-5) {};

                  \draw[-] (r) -- (v1) node[midway,xshift=6pt,yshift=0pt]{$ \delta _k $};
                  \draw[-] (v1) -- (v2) node[midway,xshift=6pt,yshift=0pt]{$ \delta _k $};
                  \draw[-] (v4) -- (c) node[midway,xshift=6pt,yshift=0pt]{$ \delta _k $};
                  \draw[-] (c) -- (w1) node[midway,xshift=-6pt,yshift=-6pt]{$ \delta _k' $};
                  \draw[-] (w1) -- (w2) node[midway,xshift=-6pt,yshift=-6pt]{$ \delta _k' $};
                  \draw[-] (w4) -- (4) node[midway,xshift=-6pt,yshift=-6pt]{$ \delta _k' $};

                  \draw[decorate,decoration = {brace,raise=15pt}] (c) -- (4) node[midway,xshift=-20pt,yshift=-20pt] {\footnotesize{$ \frac{1}{\sqrt{2}} $}};

                  \draw[decorate,decoration = {brace,raise=15pt}] (r) -- (c) node[midway,xshift=25pt,yshift=0pt] {\footnotesize{$ 2 $}};
              \end{tikzpicture}
              \caption{Illustration of the paths represented by dashed lines in Figure \ref{fig:lowerbounddeviation} by a single $ t_i $.}
              \label{fig:lowerbounddeviation2}
      \end{subfigure}
      \caption{Instance defined in the proof of Theorem~\ref{thm:lowerbounddeviation}. Superscripts are  omitted, e.g.\ $r = r^{(k)}$. Solid lines represent edges, each dashed line represents a path. Black edge labels denote edge/path lengths, blue vertex labels denote terminal weights.}
   \end{figure}

   We will construct instances with underlying graph metrics  induced by graphs indicated in  Figure~\ref{fig:lowerbounddeviation}.
   For $k\in\mathbb{N}$, we define the graph $G^{(k)} = (V^{(k)}, E^{(k)})$ by
   \begin{align*}
      V^{(k)} &= \{ r^{(k)}, c^{(k)}, t_1^{(k)}, \dots, t_k^{(k)}, v_1^{(k)}, \dots, v_q^{(k)} \} \\
      \text{and} \ E^{(k)} &= E_{r,c}^{(k)} \ \dot\cup \ E_{c,t}^{(k)} \ \dot\cup \ E_{r,t}^{(k)} \text{,}
   \end{align*}
   where   $q$ is chosen sufficiently large to provide sufficiently many inner path vertices $v_i^{(k)}$ $(1\le i\le q)$ in the following definitions of  $E_{r,c}^{(k)}$, $E_{c,t}^{(k)}$ and $\ E_{r,t}^{(k)}$:
   For $0 < \delta_k < \delta'_k < \frac{1}{k}$,  $E_{r,c}^{(k)}$ contains edges of length $\delta_k$ forming a path of total length $2$ between $r^{(k)}$ and $c^{(k)}$, and $E_{c,t}^{(k)}$ contains edges of length $\delta'_k$, forming paths of length $\frac{1}{\sqrt{2}}$ between $c^{(k)}$ and each $t_i^{(k)}$. Lastly,
   \begin{equation*}
      E_{r,t}^{(k)} = \{ \{ r^{(k)}, t_i^{(k)} \} \ | \ i = 1, \dots, k \}
   \end{equation*}
   connects each $t_i^{(k)}$ directly to $r^{(k)}$ with an edge of length $1$. The terminals are given by $T^{(k)} = V^{(k)} \setminus \{r^{(k)}\}$, and the delay weights $w^{(k)} \colon T^{(k)} \to \mathbb{R}_{\geq 0}$ are defined as
   \begin{equation*}
      w^{(k)}(t) = \begin{cases}\frac{1}{\sqrt{2}} & \text{if} \ t \in \{t_1^{(k)},\dots,t_k^{(k)}\} \\
      0 & \text{else.}\end{cases}
   \end{equation*}

   Now, the lower bound becomes:
   \begin{equation} C^{(k)} +  D^{(k)} =  2 + \frac{k}{\sqrt{2}} + \sum_{i=1}^{k} \frac{1}{\sqrt{2}} \ \text{dist}_{G^{(k)}}(r^{(k)}, t_i^{(k)}) =   2 + 2 \ \frac{k}{\sqrt{2}}  = 2 + \sqrt{2} \ k.
     \label{eq:deviationLB}
   \end{equation}

   We claim that every optimum solution contains all edges of the form $\{r^{(k)}, t_i^{(k)}\}$. Additionally, we claim that all optimum solutions contain all edges of length $\delta_k$ and all but $k$ edges of length $\delta'_k$.
   This determines the structure of an optimum solution up to the choice of the $k$ ommitted edges.
   The  length of an optimum solution is $(1 + \frac{1}{\sqrt{2}})k + 2 - \delta'_k \ k$, and its objective is
   \begin{equation}\label{eq:deviationOPT}
      \mathrm{OPT}^{(k)} = \left( 1 + \frac{1}{\sqrt{2}}\right) k + 2 - \delta'_k k + \frac{k}{\sqrt{2}} = (1+\sqrt{2}) k + 2 - \delta'_k k \text{.}
   \end{equation}

   Combining (\ref{eq:deviationLB}) and (\ref{eq:deviationOPT}), we see that
   \begin{equation*}
      \lim_{k\to\infty} \frac{\mathrm{OPT}^{(k)}}{C^{(k)} + D^{(k)}}
      = \lim_{k\to\infty} \frac{(1+\sqrt{2})k + 2 - \delta'_k k}{2 + \sqrt{2} \ k}
      = \frac{1 + \sqrt{2}}{\sqrt{2}}
      = 1 + \frac{1}{\sqrt{2}}
   \end{equation*}
   as stated in the theorem.

   To prove the first claim, assume there is an optimum solution $Y^*$ not containing an edge $\{r^{(k)},t_i^{(k)}\}$ for some $i \in \{1,\dots,k\}$. First, observe that any path from $r^{(k)}$ to $t_i^{(k)}$ not using the edge $\{r^{(k)},t_i^{(k)}\}$ contains $c$, so we have
   \begin{equation*}
      \text{dist}_{Y^*}(r^{(k)}, t_i^{(k)})
      \geq \text{dist}_{G^{(k)}}(r^{(k)},c^{(k)}) + \text{dist}_{G^{(k)}}(c^{(k)},t_i^{(k)})
      = 1 + \frac{1}{\sqrt{2}} + \frac{1}{\sqrt{2}}
      = 1 + \sqrt{2} \text{.}
   \end{equation*}

   Let $e$ be the edge of length $\delta'_k$ adjacent to $t_i^{(k)}$. Then $e\in E(Y^*)$, as otherwise $t_i^{(k)}$ would be isolated in $Y^*$. Now define $Y'$ from $Y^*$ by adding $\{r^{(k)},t_i^{(k)}\}$ and removing $e$. This increases the connection cost by $1-\delta'_k$.  The delay cost decreases by at least
   \begin{equation*}
      w(t_i^{(k)})\left( \text{dist}_{Y^*}(r^{(k)},t_i^{(k)}) - \text{dist}_{Y'}(r^{(k)},t_i^{(k)}) \right)
      \geq \frac{1 + \sqrt{2} - 1}{\sqrt{2}}
      = 1,
   \end{equation*}
   where we use $\text{dist}_{Y'}(r^{(k)},t_j^{(k)}) \leq \text{dist}_{Y^*}(r^{(k)},t_j^{(k)})$ for $j\not = i$.  Thus, the total cost  decreases  by at least $1 - (1-\delta'_k) = \delta'_k$, a contradiction to the optimality of $Y^*$.

   Now we prove the second claim: By the first claim all optimum solutions have the same delay cost $\frac{k}{\sqrt{2}}$. Hence, only the connection cost for the remaining terminals is relevant.
   From each maximal path ending in $c$ consisting only of short edges of length either $\delta_k$ or $\delta'_k$, any solution must contain either all edges or all but one. Furthermore, there must be such a path from which the solution contains all edges, otherwise there would be no $r$-$c$-path. Since $\delta_k < \delta'_k$, the shortest such configuration is to take all edges of length $\delta_k$ and all but $k$ edges of length $\delta'_k$ (namely all but one from each path).
\end{proof}

Even in the Manhattan plane this gap is still at least $\sqrt{2}$, as we show in
Appendix~\ref{sec:gap-lb-manhattan}.

\section{The \texorpdfstring{$(1+\beta)$}{(1 + β)}-approximation algorithm}
\label{sec:approximation-algorithm}

For shorter formulas, we will use the following notation in the remainder of this paper.
Let $A$ be an arborescence. By   $ A_v $ we denote the sub-arborescence rooted at $v$.
Furthermore, $ T_{A} := V(A)\cap T $ is the set of terminals in $A$,
$ W_{A} := w(T_{A}) $ is the sum of delay weights in $A$,
$ C_{A} := c(E(A)) $ is  the \textbf{connection cost} of $A$ and
$ D_{A} := D_{T_A} := \sum_{t\in T_{A}} w(t)  c(r,t) $ the \textbf{minimum possible delay cost} for connecting the sinks in  $T_{A}$ (independent of the structure of $A$).

Recall that $ \beta \geq 1 $ is the approximation guarantee for the minimum-length Steiner tree problem. The algorithm in \cite{Khazraei21} is  described in Algorithm~\ref{alg:khazraei21}.
After orienting its edges, we can consider any solution  $A$ as an $r$-arborescence.
We use arborescences instead of trees to simplify the algorithmic notation.

\begin{algorithm}[tb]
  \textbf{Step 1 (initial arborescence):}

  First,  compute a  $ \beta $-approximate minimum cost Steiner $r$-arborescence $A_0$ for $ T\cup \{ r\} $
  with outdegree 0 at all sinks in $T$ and outdegree 2 at all Steiner vertices in $ V(A_0)\setminus (T\cup \{ r\})$.

  ~

\textbf{Step 2 (split into branching):}

Traverse $A_0$  bottom-up. For each traversed edge $(x,y)\in E(A_0)$,
if  $ W_{(A_0)_y} > \mu$, remove the edge $(x,y)$  creating a new arborescence $(A_0)_y$ in the branching.

~

Let $\mathcal{A}$ denote the set of all arborescences that were cut off from $A_0$  this way.

~

\textbf{Step 3  (reconnect arborescences):}

reconnect each sub-arborescence  $A'$ that was cut off in Step 2 as follows:
Select a vertex $t \in T' := T_{A'}$ that minimizes
the cost for serving the sinks in $T'$ through the $r$-arborescence $A'+(r,t)$,
i.e.\ select a vertex $t \in T'$ as a \textbf{port} for $T'$  that minimizes
\[  c(r,t) + C_{A'}  + \sum _{t'\in T'} w(t')\cdot (c(r,t) +  c(E(A'_{[t,t']}))).\]

Let $t_1,\dots,t_{|\mathcal{A}|} \in T$ be the set of selected port vertices.
Return  the union of the final branching and
the port connections $A_0  + \{(r,t_i):  i\in \{1,\dots,|\mathcal{A}|\} \ \}$.

\caption{$(1+\beta)$-approximation algorithm by \cite{Khazraei21} using a parameter $\mu > 0$.}
\label{alg:khazraei21}
\end{algorithm}

\subsection{Essential steps for a \texorpdfstring{$ 1+\beta $}{1 + β} approximation}

We quickly recap the essential steps in the analysis of \cite{Khazraei21}, which we will use in our analysis.
The cost to connect an arborescence $A'\in\mathcal{A}$ to the root $ r $ can be estimated as follows:

\begin{lemma}[Khazraei and Held \cite{Khazraei21}, Lemma 1]
  \label{lem:kh-21}
  Let $A'  \in \mathcal{A}$  with corresponding terminal set $ T' $. By the choice of the port $ t\in T'$, the $r$-arborescence  $(A' + \{r,t\})$  has a total cost at most
  \begin{align}
        \label{eqn:kh_bound-1}
        & \phantom{\le }\; C_{A'} + \sum _{e = (x,y)\in E(A')} \frac{2 W_{A'_y} (W_{A'} - W_{A'_y})}{W_{A'}} c(e) + \left( 1 + \frac{1}{W_{A'}}\right) D_{T'} \\
        & \le \left( 1+\frac{W_{A'}}{2} \right) C_{A'} + \left( 1 + \frac{1}{W_{A'}}\right) D_{T'}         \label{eqn:kh_bound-2}\\
        & \le      \left( 1+\mu \right) C_{A'} + \left( 1 + \frac{1}{\mu}\right) D_{T'}.         \label{eqn:kh_bound-3}
  \end{align}
\end{lemma}
We sketch the proof in Appendix~\ref{sec:proof-lem-kh21}, because we use the bounds (\ref{eqn:kh_bound-1}) and (\ref{eqn:kh_bound-2})
that were not stated explicitly in \cite{Khazraei21}, Lemma 1.
While  the bounds (\ref{eqn:kh_bound-1}) and (\ref{eqn:kh_bound-2}) hold for any (sub-)arborescence $A'$,
(\ref{eqn:kh_bound-3}) depends on the specific way how  $A'\in \mathcal{A} $ was cut off during Step 2 of Algorithm~\ref{alg:khazraei21}.

A similar cost bound can  be shown easily for the arborescence $A_r$ containing the root $ r $
after Step 2. Summing up the resulting cost bounds and choosing $\mu = \frac{1}{\beta}$,
 \cite{Khazraei21} obtain the approximation factor $(1+\beta)$.

\section{Improving the approximation ratio}
\label{sec:improvement}

Algorithm~\ref{alg:khazraei21} suffers from the following weakness indicated in Figure~\ref{fig:candidate-for-further-split}.
Assume that after splitting we are given a sub-arborescence $A'\in \mathcal{A}$ with a high delay weight $W_{A'}$, a high connection cost $C_{A'}$, but a low minimum possible delay cost $D_{A'}$, e.g. as
shown in Figure~\ref{fig:candidate-for-further-split-b}.
Then Algorithm~\ref{alg:khazraei21} would retain the high delay cost.
Instead, it would be better to split the arborescence further to achieve a lower delay cost as in Figure~\ref{fig:candidate-for-further-split-c}.

In this section, we propose a refined splitting criterion that provides a better approximation ratio.
Instead of using a fixed threshold $\mu$, we allow to split off sub-arborescences earlier if their expected reconnection cost (\ref{eqn:kh_bound-1}) is sufficiently cheap.
The precise criterion is specified in  (\ref{eq:cutoffcriterion}) (inside Algorithm~\ref{alg:modification_cut-off_routine}). Observe that (\ref{eq:cutoffcriterion})
provides cheaper solutions than (\ref{eqn:kh_bound-3}), as one occurrence of  $\mu$ is replaced by $\frac{\mu}{2}$.

Then we show in Lemma~\ref{lem:mu_bound_in_Ar} that every sub-arborescence of the remaining root component has delay weight at most $\mu$.
This allows us to prove a similar improved cost bound for the root component in Lemma~\ref{lem:root-component}.

In Section~\ref{sec:main-result}, we simply combine all sub-arborescences and choose $\mu$ to prove Theorem~\ref{thm:main}.  Theorem~\ref{thm:lb_gap} follows as an immediate consequence.

\begin{figure}[bt]
        \centering
        \begin{subfigure}{0.3\textwidth}
                \centering
                \begin{tikzpicture}
                        \node[fsquare,red,label=180:$ r $] (r) at (0,0) {};

                        \node[fcircle,label=90:\footnotesize{\textcolor{blue}{$ 1 $}}] (5) at (90:1.5) {};
                        \node[fcircle,label=45:\footnotesize{\textcolor{blue}{$ 0 $}}] (4) at (45:1.5) {};
                        \node[fcircle,label=0:\footnotesize{\textcolor{blue}{$ 0 $}}] (3) at (0:1.5) {};
                        \node[fcircle,label=315:\footnotesize{\textcolor{blue}{$ 0 $}}] (2) at (315:1.5) {};
                        \node[fcircle,label=270:\footnotesize{\textcolor{blue}{$ 0 $}}] (1) at (270:1.5) {};
                        \node[fcircle,label=225:\footnotesize{\textcolor{blue}{$ 1 $}}] (0) at (225:1.5) {};

                        \draw[-] (5) to (4);
                        \draw[-] (4) to (3);
                        \draw[-] (3) to (2);
                        \draw[-] (2) to (1);
                        \draw[-] (1) to (0);

                        \draw[-] (r) to (5);
                \end{tikzpicture}
                \caption{Minimum-length Steiner tree}
                \label{fig:candidate-for-further-split-a}
        \end{subfigure}
        ~\hspace{0.025\columnwidth}
        \begin{subfigure}{0.3\textwidth}
                \centering
                \begin{tikzpicture}
                        \node[fsquare,red,label=180:$ r $] (r) at (0,0) {};

                        \node[fcircle,label=90:\footnotesize{\textcolor{blue}{$ 1 $}}] (5) at (90:1.5) {};
                        \node[fcircle,label=45:\footnotesize{\textcolor{blue}{$ 0 $}}] (4) at (45:1.5) {};
                        \node[fcircle,label=0:\footnotesize{\textcolor{blue}{$ 0 $}}] (3) at (0:1.5) {};
                        \node[fcircle,label=315:\footnotesize{\textcolor{blue}{$ 0 $}}] (2) at (315:1.5) {};
                        \node[fcircle,label=270:\footnotesize{\textcolor{blue}{$ 0 $}}] (1) at (270:1.5) {};
                        \node[fcircle,label=225:\footnotesize{\textcolor{blue}{$ 1 $}}] (0) at (225:1.5) {};

                        \draw[->] (5) to (4);
                        \draw[->] (4) to (3);
                        \draw[->] (3) to (2);
                        \draw[->] (2) to (1);
                        \draw[->] (1) to (0);

                        \draw[dashed,->] (r) to (5);
                \end{tikzpicture}
                \caption{Cost: $ 6 + (1 + 6) = 13 $}
                \label{fig:candidate-for-further-split-b}
        \end{subfigure}
        ~\hspace{0.025\columnwidth}
        \begin{subfigure}{0.3\textwidth}
                \centering
                \begin{tikzpicture}
                        \node[fsquare,red,label=180:$ r $] (r) at (0,0) {};

                        \node[fcircle,label=90:\footnotesize{\textcolor{blue}{$ 1 $}}] (5) at (90:1.5) {};
                        \node[fcircle,label=45:\footnotesize{\textcolor{blue}{$ 0 $}}] (4) at (45:1.5) {};
                        \node[fcircle,label=0:\footnotesize{\textcolor{blue}{$ 0 $}}] (3) at (0:1.5) {};
                        \node[fcircle,label=315:\footnotesize{\textcolor{blue}{$ 0 $}}] (2) at (315:1.5) {};
                        \node[fcircle,label=270:\footnotesize{\textcolor{blue}{$ 0 $}}] (1) at (270:1.5) {};
                        \node[fcircle,label=225:\footnotesize{\textcolor{blue}{$ 1 $}}] (0) at (225:1.5) {};

                        \draw[->] (4) to (3);
                        \draw[->] (3) to (2);
                        \draw[->] (2) to (1);
                        \draw[->] (5) to (4);

                        \draw[dashed,->] (r) to (5);
                        \draw[dashed,->] (r) to (0);
                \end{tikzpicture}
                \caption{Cost: $ 6 + (1 + 1) = 8 $}
                \label{fig:candidate-for-further-split-c}
        \end{subfigure}
        \caption{Weakness of Algorithm~\ref{alg:khazraei21}: $ (M,c) $ is induced by a complete graph with seven vertices and unit weights. Delay weights are indicated by the blue node labels and $\mu = 1$. Algorithm~\ref{alg:khazraei21} might start with the minimum-length Steiner tree on the left. Then the algorithm will cut the edge incident to $ r $ and reconnect the sub-arborescence resulting possibly in the solution in the middle. On the right the result from our improved algorithm is shown.}
        \label{fig:candidate-for-further-split}
\end{figure}

\subsection{Improving the splitting routine}
\label{sec:cut-off routine}

Algorithm~\ref{alg:modification_cut-off_routine} shows our improved splitting step, which cuts off a sub-arborescence
if we can reconnect it cheaply, i.e.\ if  (\ref{eq:cutoffcriterion}) holds.
\begin{algorithm}[tb]
        \textbf{Step 2 (split into branching):}

        Traverse $A_0$  bottom-up. For each traversed edge $(v,z)\in E(A_0)$ consider $ A_z := (A_0)_z $: If $ W_{A_z} > 0 $ and
                \begin{equation}\label{eq:cutoffcriterion}
                   \sum _{e = (p,q)\in E(A_z)} \frac{2 W_{(A_z)_q} (W_{A_z} - W_{(A_z)_q})}{W_{A_z}} c(e) + \frac{D_{A_z}}{W_{A_z}}
                   \leq \frac{\mu}{2} \left( C_{A_z} + c(v,z)\right) + \frac{D_{A_z}}{\mu},
                \end{equation}
                remove $ (v,z) $ creating a new arborescence $ A_z $.

\caption{Modifying Step 2 of Algorithm \ref{alg:khazraei21}}
\label{alg:modification_cut-off_routine}
\end{algorithm}
With Lemma \ref{lem:kh-21} we immediately get the following result for the cut-off sub-arborescences:
\begin{lemma}
        \label{lem:improved_cut-off_routine}
        Let $ A'\in \mathcal{A} $ be an arborescence that was cut off in  Algorithm~\ref{alg:modification_cut-off_routine} and let $ e_{A'} $ be the incoming edge in the root of the arborescence $ A' $ which was deleted during this step. Then the corresponding terminals in $ T_{A'} $ can be connected to the root $ r $ with total cost at most
        \begin{align*}
                \left( 1 + \frac{\mu}{2} \right) \left( C_{A'} + c(e_{A'})\right) + \left( 1 + \frac{1}{\mu}\right) D_{A'}.
        \end{align*}\qed
\end{lemma}

After the original Step 2 of  Algorithm \ref{alg:khazraei21}, it is clear that for all edges $ (r,x)\in \delta _{A_0}^+ (r) $ of the root component the total delay  weight $ W_{(A_0)_x} $ is at most $ \mu $.
We show that this also holds after the modified Step 2 in Algorithm~\ref{alg:modification_cut-off_routine}. However, the analysis is more complicated and uses the following two functions.
\begin{definition}
  \label{def:f-and-g}
        Let $ \mu > 0 $ and $X^{\mu} := \{ (a,b,c) \in (\mu ,2\mu )\times (0,\mu)^2: c\leq a-b < \mu \}$.
        We define the functions $ f, g\colon X^\mu \to \mathbb{R} $ as
    \begin{align*}
                f (a,b,c)
                &:= \frac{2(a-c) c}{a} - \frac{\mu}{2} + \left( \frac{1}{a} - \frac{1}{\mu}\right) \cdot \frac{1}{\frac{1}{a-b} - \frac{1}{\mu}}\cdot \left( \frac{\mu}{2} - \frac{2 ((a-b)-c) c}{a-b}\right) \\
                g (a,b,c)
                &:= \frac{2(a-c) c}{a} - \frac{\mu}{2} + \left( \frac{1}{a} - \frac{1}{\mu}\right) \cdot \frac{1}{\frac{1}{a-b} - \frac{1}{\mu}}\cdot \frac{\mu}{2} .
        \end{align*}
\end{definition}

\begin{lemma}
  \label{lem:non-positive_functions}
  For all $ (a,b,c)\in X^\mu $,  $ f(a,b,c)\leq 0 $ and $ g(a,b,c)\leq 0$.
\end{lemma}
A proof of Lemma~\ref{lem:non-positive_functions} based on algebraic transformations can be found in Appendix~\ref{sec:proof-lem-non-positive_functions}.

\begin{lemma}
  \label{lem:mu_bound_in_Ar}
  After cutting off sub-arborescences with Algorithm~\ref{alg:modification_cut-off_routine},
  every child $ x\in \Gamma _{A_r}^+ (r) $ of $r$ in the remaining root component $ A_r := (A_0)_r $ satisfies $ W_{(A_r)_x}\leq \mu $.
\end{lemma}

\begin{proof}
        Assume the opposite would be true. Let $ z $ be a vertex in $ A_r - r $ such that the weight of the sub-arborescence $ A_z := (A_r)_z $ exceeds $ \mu $ and the weight of every child arborescence $ (A_z)_x $ is at most $ \mu $ for all edges $ (z,x)\in \delta _{A_z}^+ (z) $. We distinguish  two cases:

        \textbf{Case 1.} $ z $ is a terminal. Then $ z $ is also a leaf and the left-hand side of (\ref{eq:cutoffcriterion}) simplifies to
        \begin{align*}
               \frac{1}{W_{A_z}} D_{A_z}
               \leq \frac{1}{\mu} D_{A_z}
        \end{align*}
        since $ A_z $ does not contain any edges. But then $ A_z $ would have  been cut-off in Step 2, a contradiction.

        \textbf{Case 2.} $ z $ is a Steiner vertex. Then $ z $ has two outgoing edges $ e_x := (z,x), e_y :=(z,y)\in \delta _{A_z}^+ (z)$ as shown in Figure~\ref{fig:setting-nonterminal-z}. A single outgoing edge would contradict the choice of $ z $. With $ A_x := (A_z)_x $ or $ A_y := (A_z)_y $ this implies $ 0 < W_{A_x}, W_{A_y}\leq \mu $. If $ W_{A_x} = \mu $,  Lemma \ref{lem:kh-21}, (\ref{eqn:kh_bound-2}) shows that $ A_x $ satisfied the bound (\ref{eq:cutoffcriterion}) when it was considered in Step 2 and would have been cut off. Analogously,  $ W_{A_y} \ne \mu$.
                \begin{figure}
                \centering
                \begin{tikzpicture}[scale=1.05]
                        \node[hcircle,label=above:$ z $] (v) at (0,0) {};
                        \node[hcircle,label=left:$ x $] (x) at (-1.25,-2) {};
                        \node[hcircle,label=right:$ y $] (y) at (1.25,-2) {};

                        \draw[->] (v) to node[above,sloped] {$ e_x $} (x);
                        \draw[->] (v) to node[above,sloped] {$ e_y $} (y);

                        \draw (x) to (-2,-4) to (-0.5,-4) to (x);
                        \node at (-1.25,-3.25) {$ A_x $};

                        \draw (y) to (0.5,-4) to (2,-4) to (y);
                        \node at (1.25,-3.25) {$ A_y $};
                \end{tikzpicture}
                \caption{Setting in the proof of Lemma~\ref{lem:mu_bound_in_Ar} if $z$ is  a Steiner vertex (Case 2).}
                \label{fig:setting-nonterminal-z}
        \end{figure}
        Thus, $ W_{A_x}, W_{A_y} < \mu $. Since (\ref{eq:cutoffcriterion}) does not hold for $A_x$, we get (by transforming its negation)
\begin{align*}
        \underbrace{\left( \frac{1}{W_{A_x}} - \frac{1}{\mu}\right)}_{> 0} D_{A_x} >
        \sum _{e = (u,v)\in E(A_x)} \left(\frac{\mu}{2} - \frac{2 (W_{A_x} - W_{(A_x)_{v}}) W_{(A_x)_{v}}}{W_{A_x}}\right) c(e) + \frac{\mu}{2} c(e_x).
\end{align*}
Combining this with  the analogue inequality for $A_y$ and using $D_{A_z}=D_{A_x}+D_{A_y}$, we get

\begin{align*}
   &\underbrace{\left( \frac{1}{W_{A_z}} - \frac{1}{\mu}\right)}_{< 0} D_{A_z} \\
   &\quad < \left( \frac{1}{W_{A_z}} - \frac{1}{\mu}\right) \bigg( \sum _{e = (u,v)\in E(A_x)} \frac{1}{\frac{1}{W_{A_x}} - \frac{1}{\mu}} \left(\frac{\mu}{2} - \frac{2 (W_{A_x} - W_{(A_x)_{v}}) W_{(A_x)_{v}}}{W_{A_x}}\right) c(e) \\
   &\quad\hphantom{\left( \frac{1}{W_{A_z}} - \frac{1}{\mu}\right) \bigg(} \! + \frac{\frac{\mu}{2}}{\frac{1}{W_{A_x}} - \frac{1}{\mu}} c(e_x) \\
   &\quad\hphantom{\left( \frac{1}{W_{A_z}} - \frac{1}{\mu}\right) \bigg(} \! + \sum _{e = (u,v)\in E(A_y)} \frac{1}{\frac{1}{W_{A_y}} - \frac{1}{\mu}} \left(\frac{\mu}{2} - \frac{2 (W_{A_y} - W_{(A_y)_{v}}) W_{(A_y)_{v}}}{W_{A_y}}\right) c(e) \\
   &\quad\hphantom{\left( \frac{1}{W_{A_z}} - \frac{1}{\mu}\right) \bigg(} \! + \frac{\frac{\mu}{2}}{\frac{1}{W_{A_y}} - \frac{1}{\mu}} c(e_y) \bigg).
\end{align*}
This inequality together with
\begin{align*}
   &\sum _{e = (u,v)\in E(A_z)} \left( \frac{2 (W_{A_z} - W_{(A_z)_{v}}) W_{(A_z)_{v}}}{W_{A_z}} - \frac{\mu}{2}\right) c(e) \\
   &= \sum _{e = (u,v)\in E(A_x)} \left( \frac{2 (W_{A_z} - W_{(A_z)_{v}}) W_{(A_z)_{v}}}{W_{A_z}} - \frac{\mu}{2}\right) c(e) \\
   &\quad + \left( \frac{2 (W_{A_z} - W_{(A_z)_x}) W_{(A_z)_x}}{W_{A_z}} - \frac{\mu}{2}\right) c(e_x) \\
   &\quad + \sum _{e = (u,v)\in E(A_y)} \left( \frac{2 (W_{A_z} - W_{(A_z)_{v}}) W_{(A_z)_{v}}}{W_{A_z}} - \frac{\mu}{2}\right) c(e) \\
   &\quad + \left( \frac{2 (W_{A_z} - W_{(A_z)_y}) W_{(A_z)_y}}{W_{A_z}} - \frac{\mu}{2}\right) c(e_y)
\end{align*}
yields
\begin{align*}
        &\sum _{e = (u,v)\in E(A_z)} \left( \frac{2 (W_{A_z} - W_{(A_z)_{v}}) W_{(A_z)_{v}}}{W_{A_z}} - \frac{\mu}{2}\right) c(e) + \left( \frac{1}{W_{A_z}} - \frac{1}{\mu}\right) D_{A_z} \\
        &\quad < \sum _{e = (u,v)\in E(A_x)} f (W_{A_z},W_{A_y},W_{(A_z)_{v}}) c(e) + \sum _{e = (u,v)\in E(A_y)} f (W_{A_z},W_{A_x},W_{(A_z)_{v}}) c(e) \\
        &\quad\quad + g(W_{A_z}, W_{A_y}, W_{(A_z)_{v}}) c(e_x) + g(W_{A_z}, W_{A_x}, W_{(A_z)_{v}}) c(e_y).
\end{align*}
By  Lemma \ref{lem:non-positive_functions} and $ 0 < W_{A_x}, W_{A_y} < \mu $, the last term is non-positive. Therefore $ A_z $ satisfied the bound (\ref{eq:cutoffcriterion}) when it was considered in Step 2 and would have been cut off, a contradiction.
\end{proof}

In \cite{Khazraei21} the   final root arborescence $ A_r$, which   was not cut off in Step 2 of Algorithm~\ref{alg:khazraei21}, was kept unaltered.
Using  Lemma~\ref{lem:mu_bound_in_Ar}, we  show how to connect it  in a better way.

\begin{lemma}
        \label{lem:root-component}
        Let $ A_r $ be the sub-arborescence of $ A_0 $ rooted at $ r $ after the modified Step 2  of Algorithm \ref{alg:khazraei21}.
        The terminal set $ T_{A_r} $ can be connected to the root $ r $ with total cost at most
        \begin{align*}
                \left( 1 + \frac{\mu}{2} \right) C_{A_r} + \left( 1 + \frac{1}{\mu}\right) D_{A_r}.
        \end{align*}
\end{lemma}
\begin{proof}
        Let $ (r,x)\in \delta _{A_r}^+ (r) $ be arbitrary and $ A_x $ the arborescence of $ A_r-r $ rooted at $ x $. We show that the terminal set $ T_{A_x} $ can be connected to the root $ r $ with total cost at most
        \begin{align*}
                \left( 1 + \frac{\mu}{2}\right) (C_{A_x} + c(r,x)) + \left( 1 + \frac{1}{\mu}\right) D_{A_x}.
        \end{align*}
                Adding this cost for all edges in $ \delta _{A_r}^+ (r) $, we obtain the claim.

        We distinguish between two cases:

        \textbf{Case 1.}
              \[
                        W_{A_x} (C_{A_x} + c(r,x))
                        \leq \frac{\mu}{2} (C_{A_x} + c(r,x)) + \frac{1}{\mu} D_{A_x}.
                \]
                By keeping the arborescence $ A_x $ connected through $ (r,x) $, the connection cost is $ C_{A_x} + c(r,x) $. In particular, for each terminal $ t\in T_{A_x} $, the $ r $-$ t $-path in $ A_x + (r,x) $ has a length of at most $ C_{A_x} + c(r,x) $. We therefore obtain a total cost of at most
                \[
                        (1 + W_{A_x}) (C_{A_x} + c(r,x))
                        \leq \left( 1 + \frac{\mu}{2} \right) (C_{A_x} + c(r,x)) + \frac{1}{\mu} D_{A_x} .
                \]

         \textbf{Case 2.}
                \begin{align}
                        \label{eq91x}
                        W_{A_x} (C_{A_x} + c(r,x))
                        > \frac{\mu}{2} (C_{A_x} + c(r,x)) + \frac{1}{\mu} D_{A_x}.
                \end{align}
                Therefore we have $ W_{A_x} > 0 $ and obtain from (\ref{eq91x}) an upper bound on the minimum possible delay cost of $ A_x $
                \begin{align}
                        \label{eq90x}
                        D_{A_x}
                        < \left( W_{A_x} - \frac{\mu}{2}\right) \mu (C_{A_x} + c(r,x)).
                \end{align}
                We remove the edge $ (r,x) $ and connect the arborescence $ A_x $ to the root $ r $. By Lemma~\ref{lem:mu_bound_in_Ar}, $W_{A_x} \le \mu$. As in Lemma \ref{lem:kh-21} we obtain total cost of at most
                \begin{align*}
                        &\hspace{-1em}\left( 1 + \frac{W_{A_x}}{2}\right) C_{A_x} + \left( 1 + \frac{1}{W_{A_x}}\right) D_{A_x} \\
                        &= \left( 1 + \frac{W_{A_x}}{2}\right) C_{A_x} + \left( 1 + \frac{1}{\mu}\right) D_{A_x} + \underbrace{\frac{\mu - W_{A_x}}{\mu W_{A_x}}}_{\geq 0} D_{A_x} \\
                        \overset{(\ref{eq90x})}&{\leq} \left( 1 + \frac{W_{A_x}}{2}\right) C_{A_x} + \left( 1 + \frac{1}{\mu}\right) D_{A_x} + \frac{\mu - W_{A_x}}{\mu W_{A_x}} \left( W_{A_x} - \frac{\mu}{2}\right) \mu (C_{A_x} + c(r,x)) \\
                        &\leq \left( 1 - \frac{W_{A_x}}{2} + \frac{3}{2} \mu - \frac{\mu ^2}{2 W_{A_x}}\right) (C_{A_x} + c(r,x)) + \left( 1 + \frac{1}{\mu}\right) D_{A_x}.
                \end{align*}
                With the following estimation we obtain the claimed bound
                \begin{align*}
                        -\frac{W_{A_x}}{2} + \frac{3}{2} \mu - \frac{\mu ^2}{2 W_{A_x}}
                        = \frac{\mu}{2} - \frac{1}{2} \left( \sqrt{W_{A_x}} - \frac{\mu}{\sqrt{W_{A_x}}}\right) ^2
                        \leq \frac{\mu}{2}.
                \end{align*}
\end{proof}

\begin{theorem}\label{thm:running_time}
   Algorithm~\ref{alg:modification_cut-off_routine} and the reconnect in Step 3  as well as of the root component can be implemented to run in time $\mathcal{O}(|T|)$.
\end{theorem}
\begin{proof}
  A na\"ive implementation would immediately result in a quadratic running time.
  We can achieve a linear running time by computing all relevant information  incrementally in constant time per node
  during the bottom-up traversal. Details can be found in Appendix~\ref{appendix:running-time}.
\end{proof}

\subsection{Proving Theorem~\ref{thm:main} and Theorem~\ref{thm:lb_gap}}
\label{sec:main-result}

We start by analyzing the combination of all sub-arborescences.
\begin{theorem}
        \label{thm:main-for-given-mu}
        Given an instance $(T,r,w)$ of the \textsc{Uniform Cost-Distance Steiner Tree Problem}, we can compute in $ \mathcal{O} (\Lambda + |T|) $ time a Steiner tree with objective value at most
        \begin{align}
                \label{eq96}
                \left( 1 + \frac{\mu}{2} \right) C + \left( 1 + \frac{1}{\mu}\right) D,
        \end{align}
        where $ C  $ is the cost of a $ \beta $-approximate minimum-length Steiner tree and $D := D(T,r,w)$. Here, $ \Lambda $ is the running time for computing a $ \beta $-approximate minimum Steiner tree for $ T\cup \{ r\} $.
\end{theorem}
\begin{proof}
We run Algorithm~\ref{alg:khazraei21} with two modifications:
\begin{enumerate}
\item The cut-off routine (Step 2) is modified according to Algorithm~\ref{alg:modification_cut-off_routine}.

\item The arborescence $A_r$ containing the  root $ r $ after Step 2 is  reconnected to the root $ r $ according to Lemma~\ref{lem:root-component}.
\end{enumerate}

The total cost of the computed solution is  upper bounded  by the sum of the cost bounds for these $r$-arborescences, which is (\ref{eq96}).
For the running time analysis, we consider the individual steps of the  algorithm:

In Step 1, a $ \beta $-approximate minimum Steiner tree for $ T\cup \{ r\} $ is computed in time $ \mathcal{O} (\Lambda ) $ and transformed into the arborescence $ A_0 $ obeying the degree constraints in linear time as in \cite{Khazraei21}.
The linear running time of Step 2 and Step 3  follows from Theorem~\ref{thm:running_time}.
\end{proof}

Finally, we choose the threshold $ \mu $ based on the quantities $ C $ and $ D $ to prove Theorem~\ref{thm:main}:
\begin{proof}(of Theorem~\ref{thm:main})
  We make the following modification of the algorithm in Theorem~\ref{thm:main-for-given-mu}:

  If  $ C = c(E(A_0)) = 0$, each $ r $-$ t $-path, $ t \in T $,  has length $ 0 $ in $A_0$. So this is already an optimal solution and we  just return $A_0$.

      Otherwise, set $ \mu := \sqrt{\frac{2D}{C}} $ and the  algorithm from Theorem~\ref{thm:main-for-given-mu} provides us with a solution with total cost at most
      \begin{align*}
                C + D + \sqrt{2}\sqrt{CD}
                \leq \beta C_{SMT}(T\cup\{r\}) + D + \sqrt{2}\sqrt{\beta  C_{SMT}(T\cup\{r\}) \cdot D}.
        \end{align*}
        We divide this by the lower bound       $ C_{SMT}(T\cup\{r\}) + D$ in (\ref{eq:lower_bound}). Now, the approximation factor is at most the maximum of the function $ h\colon \mathbb{R} _{> 0}\times \mathbb{R} _{\geq 0} \to \mathbb{R} $ given by
        \begin{align*}
                h(x,y)
                := \frac{\beta x + y + \sqrt{2}\sqrt{\beta x y}}{x + y} .
        \end{align*}
         By our assertion, $C_{SMT}(T\cup\{r\}) \ge \frac{C}{\beta}> 0$. In Appendix~\ref{appendix:bounding-h} we prove for $x+y>0$ using algebraic reformulations
        \begin{align*}
                h(x,y)
                \leq \beta + \frac{\beta}{\sqrt{\beta ^2 + 1} + \beta - 1},
        \end{align*}
        proving the claimed approximation ratio.
\end{proof}

Using Theorem~\ref{thm:main} we obtain the  approximation factors shown in Table~\ref{table:approx-factors} (rounded to five decimal digits) for some interesting values of $ \beta $.

\begin{table}[t]
        \centering

        \begin{tabular}{l cccc}
                \toprule
                Parameter $ \beta $ & $ 1 $ & $ \ln (4) + \epsilon $ & $ \frac{3}{2} $ & $ 2 $ \\
                \midrule
                Algorithm~\ref{alg:khazraei21} \cite{Khazraei21}& $ 2.00000 $ & $ 2.38630 $ & $ 2.50000 $ & $ 3.00000 $ \\
                 Theorem~\ref{thm:main} & $ 1.70711 $ & $ 2.04782 $ & $ 2.15139 $ & $ 2.61804 $ \\
                \bottomrule
        \end{tabular}
                \caption{Comparison of approximation factors for the \textsc{Uniform Cost-Distance Steiner Tree Problem} with different approximation factors $ \beta $ for the minimum-length Steiner tree problem.}
        \label{table:approx-factors}
\end{table}

\begin{proof} (of Theorem~\ref{thm:lb_gap})
This is a direct consequence of   Theorem~\ref{thm:lowerbounddeviation} and  Theorem~\ref{thm:main} for $\beta=1$.
\end{proof}

\subsection{Tightness of the Analysis}
\label{sec:tightness_of_analysis}
We present a family of instances where our algorithm returns
solutions that are asymptotically a factor $1+\frac{1}{\sqrt{2}}$ above the optimum, even when starting with a minimum-length Steiner tree.
For $k\in \mathbb{N}$, we are given a root $r$ and  $2k$  terminals  $T= \{u_i,v_i : 1\le i\le k\}$
that are placed on a single  line in the order $r < v_1 < u_1 <\dots < v_k < u_k$ as shown in Figure~\ref{fig:tightness-of-algorithm}  for $k=3$.

We specify the distances between adjacent terminals.
Let $u_0 := r$. The distances are
$c(u_{i-1},v_i) := \frac{1}{k}$ for $1\le i \le k$,  $c(v_1,u_1) = \frac{1}{k}$, and
\begin{equation*}
  c(v_i,u_i) = \frac{i-\sqrt{2}}{{\sqrt2}k} + \frac{1}{\sqrt{2}}\sum_{j=1}^{i-1}c(v_j,u_j) \quad \text{ for } 2 \le i \le k.
  \end{equation*}

\begin{figure}
  \centering
    \begin{tikzpicture}
      \node[fsquare,red,label=$r$] (w0) at (0,0) {};
      \node[fsquare,blue,label=$v_1$] (v1) at (1.0,0) {};
      \node[fsquare,darkgreen,label=$u_1$] (w1) at (2.0,0) {};

      \node[fsquare,blue,label=$v_2$] (v2) at (3.0,0) {};
      \node[fsquare,darkgreen,label=$u_2$] (w2) at (4.1213,0) {};

      \node[fsquare,blue,label=$v_3$] (v3) at (5.1213,0) {};
      \node[fsquare,darkgreen,label=$u_3$] (w3) at (7.74,0) {};

      \draw[-] (r) -- node[midway,below]{$1/3$} (v1) -- node[midway,below]{$1/3$} (w1) -- node[midway,below]{$1/3$}(v2) -- node[midway,below]{$\approx 0.37$}(w2) -- node[midway,below]{$1/3$}(v3) -- node[midway,below]{$\approx0.87$} (w3);
    \end{tikzpicture}
    \caption{Example of the instance demonstrating the tightness of the analysis for $k=3$.}
    \label{fig:tightness-of-algorithm}

\end{figure}

Vertex weights are $w(v_1)=2, w(v_i) = \frac{1}{\sqrt{2}}$ for $ 2\le i\le k$, and $w(u_i)=0$ for  $1 \le i \le k$.

Observe that the length of a minimum Steiner tree is
\begin{align}
  C_{SMT} & = \sum_{i=1}^k\left(c(u_{i-1}, v_i)+c(v_i,u_i)\right) \\
  & = c(u_{0}, v_1)+c(v_1,u_1) + \sum_{i=2}^k\left(c(u_{i-1}, v_i)+c(v_i,u_i)\right) \\
  & = \frac{2}{k}+ \sum_{i=2}^k\left(\frac{1}{k} +\frac{i-\sqrt{2}}{{\sqrt2}k} + \frac{1}{\sqrt{2}}\sum_{j=1}^{i-1}c(v_j,u_j)\right)\\
  & =  \frac{2}{k}+ \sum_{i=2}^k\frac{1}{\sqrt{2}} \cdot\left(\frac{i}{k} + \sum_{j=1}^{i-1}c(v_j,u_j)\right) \label{eqn:tight-counter-example-length-3}\\
  & = w(v_1)\cdot c(u_0,v_1)+ \sum_{i=2}^kw(v_i)\cdot\left(c(u_{0},v_{1})+\sum_{j=1}^{i-1}\left(c(u_{j},v_{j+1}) +  c(v_j,u_j)\right)\right)\\
  & =  D(T,r,w).
\end{align}

In this tree, which is actually a path, every terminal  has the minimum possible distance from $r$. Thus, it is an optimum solution of the
uniform cost-distance Steiner tree problem with value $2\cdot C_{SMT}$.

According to the proof of Theorem~\ref{thm:main}, the algorithm chooses $\mu=\sqrt{\frac{2D(T,r,w)}{C_{SMT}}}= \sqrt{2}$.
Thus, edges entering some $u_i$ ($i\in [k]$) will never be deleted, as $w(u_i)=0$.
Now inductively, for each edge entering a vertex $v_i$ ($i=2,\ldots,k$) in bottom up order,
the left and right side of the deletion criterion (\ref{eq:cutoffcriterion}) are both identical to
$\frac{i}{k}+\sum_{j=1}^{i-1} c(v_j,u_j)$. thus the edge $(u_{i-1},v_i)$  will be deleted.
To see this, observe that the first summand of the  left side is zero as $w(u_i)=0$, and its second summand reduces the to length of the $r$-$v_i$ path.
The right side is
\begin{align*}
  &  \frac{\mu}{2} \left( C_{A_{v_i}} + c(u_{i-1},v_i)\right) + \frac{D_{A_{v_i}}}{\mu} \\
  = & \frac{1}{\sqrt{2}} \left(c(v_i,u_i) + c(u_{i-1},v_i)\right)  +\frac{1}{\sqrt{2}}\cdot \frac{1}{\sqrt{2}}  \left(c(u_{i-1},v_i) + \sum_{j=1}^{i-1}\left(c(u_{j-1},v_j)+ c(v_j,u_j)\right)\right)\\
  = &  \frac{1}{\sqrt{2}} \left(\frac{i-\sqrt{2}}{{\sqrt2}k} + \frac{1}{\sqrt{2}}\sum_{j=1}^{i-1}c(v_j,u_j)  + \frac{1}{k} + \frac{i}{k\sqrt{2}}  + \frac{1}{\sqrt{2}}\sum_{j=1}^{i-1}c(v_j,u_j)\right)\\
  = &  \frac{i}{k} + \sum_{j=1}^{i-1}c(v_j,u_j).
\end{align*}
In a similar computation we see that the deletion criterion also holds for the case $ i=1 $, which can be omitted as the component is reconnected with the deleted edge $ \{u_0,v_1\} $ and therefore not changing the result.
Thus, the algorithm will remove all edges $\{u_{i-1},v_i\}$ $(i\in \{1,\dots,k\})$.
The cost of the resulting solution is the sum of  $C_{SMT}$, $D(T,r,w)$ ($=C_{SMT}$) and the additional connection cost
for replacing the edges $(u_{i-1},v_i)$  by $r$-$v_i$-paths:
$$
2\cdot   C_{SMT} + \sum_{i=2}^k\sum_{j=1}^{i-1}(c(u_{j-1},v_j) + c(v_j,u_j)).
$$

The  deviation factor from the optimum solution is
\begin{equation*}
  \begin{array}{rll}
     &   \frac{\displaystyle 2\cdot C_{SMT}+ \sum_{i=2}^k\sum_{j=1}^{i-1}(c(u_{j-1},v_j) + c(v_j,u_j))}{\displaystyle 2\cdot C_{SMT}}
      \\[1em]
  = &  1 + \frac{\displaystyle\sum_{i=2}^k\sum_{j=1}^{i-1}(c(u_{j-1},v_j) + c(v_j,u_j))}{\displaystyle2\left(\frac{2}{k}+ \sum_{i=2}^k\frac{1}{\sqrt{2}} \cdot\left(\frac{i}{k} + \sum_{j=1}^{i-1}c(v_j,u_j)\right)\right) } \\[3em]
    = &  \displaystyle 1 + \frac{\displaystyle\sum_{i=2}^k\sum_{j=1}^{i-1}(c(u_{j-1},v_j) + c(v_j,u_j))}{\displaystyle2\left(\frac{2}{k}+ \frac{1}{\sqrt{2}} + \frac{1}{\sqrt{2}} \sum_{i=2}^k \sum_{j=1}^{i-1}\left(c(u_{j-1},v_j)+c(v_j,u_j)\right)\right) }
\displaystyle   \xrightarrow{k\rightarrow\infty}  1+ \frac{1}{\sqrt{2}},
  \end{array}
\end{equation*}
where we substituted  $C_{SMT}$ by  (\ref{eqn:tight-counter-example-length-3}) in the first equation and used
$$
  \sum_{i=2}^k\sum_{j=1}^{i-1}(c(u_{j-1},v_j) + c(v_j,u_j))
  \geq \sum _{i=2}^k \frac{i-1}{k}
  = \frac{1}{k} \sum _{i=1}^{k-1} i
  = \frac{k-1}{2}
  \overset{k\to\infty}{\longrightarrow} \infty .
$$

\section{Worst-Case Example for any Cut-and-Reconnect Algorithm}
\label{sec:worst-case-cut-reconnect}
Any algorithm for the cost-distance problem that takes a minimum-length Steiner tree, cuts it into several pieces by deleting edges and then reconnects each piece directly to the root can at best achieve an approximation factor of $\frac{3}{2}$, as the following theorem shows.
\begin{theorem}\label{thm:worstcaseexample}
   Consider any algorithm for the cost-distance problem that works by sub-dividing a minimum-length Steiner tree into components and then reconnects the components with edges from the root directly to one vertex in each component. Then this algorithm does not have an approximation ratio better than $\frac{3}{2}$.
\end{theorem}
\begin{proof}
   \begin{figure}
      \centering
      \begin{subfigure}{0.45\textwidth}
         \begin{tikzpicture}[scale=0.9]
            \node[fsquare,red,label=$r$] (r) at (0,0) {};
            \foreach \x in {1,...,6} {
               \node[hcircle] (groundline-\x) at (\x,0) {};
               \ifthenelse{\x=1}{
                  \draw[-] (r) -- (groundline-\x);
               }{
                  \tikzmath{\l = \x - 1;}
                  \draw[-] (groundline-\l) -- (groundline-\x);
               }
            }
            \foreach \x in {1,2,3} {
               \node[fcircle,label=$t_{\x}$] (t\x) at (\x,\x) {};
            }

            \node[hcircle] (b1_1) at (2,1) {};
            \draw[-] (t1) -- (b1_1) -- (groundline-2);

            \node[hcircle] (b2_1) at (3,2) {};
            \node[hcircle] (b2_2) at (4,2) {};
            \node[hcircle] (b2_3) at (4,1) {};
            \draw[-] (t2) -- (b2_1) -- (b2_2) -- (b2_3) -- (groundline-4);

            \node[hcircle] (b3_1) at (4,3) {};
            \node[hcircle] (b3_2) at (5,3) {};
            \node[hcircle] (b3_3) at (6,3) {};
            \node[hcircle] (b3_4) at (6,2) {};
            \node[hcircle] (b3_5) at (6,1) {};
            \draw[-] (t3) -- (b3_1) -- (b3_2) -- (b3_3) -- (b3_4) -- (b3_5) -- (groundline-6);
         \end{tikzpicture}
         \caption{Vertices connected by a minimum-length Steiner tree.}
         \label{fig:worstcaseexample}
      \end{subfigure}\hspace*{30pt}\begin{subfigure}{0.45\textwidth}
         \begin{tikzpicture}[scale=0.9]
            \node[fsquare,red,label=$r$] (r) at (0,0) {};
            \foreach \x in {1,...,6} {
               \node[hcircle] (groundline-\x) at (\x,0) {};
               \ifthenelse{\x=1}{
                  \draw[-] (r) -- (groundline-\x);
               }{
                  \tikzmath{\l = \x - 1;}
                  \draw[-] (groundline-\l) -- (groundline-\x);
               }
            }
            \foreach \x in {1,2,3} {
               \node[fcircle,label=$t_{\x}$] (t\x) at (\x,\x) {};
            }

            \node[hcircle] (b1_1) at (2,1) {};
            \draw[-] (groundline-1) -- (t1) -- (b1_1);

            \node[hcircle] (b2_1) at (3,2) {};
            \node[hcircle] (b2_2) at (4,2) {};
            \node[hcircle] (b2_3) at (4,1) {};
            \draw[-] (b1_1) -- (t2) -- (b2_1) -- (b2_2) -- (b2_3);

            \node[hcircle] (b3_1) at (4,3) {};
            \node[hcircle] (b3_2) at (5,3) {};
            \node[hcircle] (b3_3) at (6,3) {};
            \node[hcircle] (b3_4) at (6,2) {};
            \node[hcircle] (b3_5) at (6,1) {};
            \draw[-] (b2_1) -- (t3) -- (b3_1) -- (b3_2) -- (b3_3) -- (b3_4) -- (b3_5);
         \end{tikzpicture}
         \caption{An optimum solution.}
         \label{fig:worstcaseexample_opt}
      \end{subfigure}
      \caption{The instance in $(\mathbb{R}^2, \ell_1)$ from the proof of Theorem~\ref{thm:worstcaseexample} for $k=3$.}
   \end{figure}

   For $k \in \mathbb{N}$ we define following instance in $(\mathbb{R}^2, \ell_1)$ (see Figure~\ref{fig:worstcaseexample}).
   Let $r = (0,0)$ and
   \begin{align*}
      T &=  \{(i,0) ~|~ i=1,\dots,2k\} \cup \bigcup _{i=1}^k B_i\text{, where}\\
      B_i &= \{t_i = (i,i)\} \cup \{(i+j,i) ~|~ j=1,\dots,i\} \cup \{(2i,j) ~|~ j=1,\dots,i-1\}.
   \end{align*}
   We define  delay weights as  $w(t_i) = \frac{2i - 1}{2i}$ for $t_i=(i,i)$ and $w(t) = 0$ elsewhere.

   Consider the following edge set of a Steiner tree for $V:= \{r\}\cup T$.
   \begin{align*}
      E &= \{ \{(i-1,0), (i,0)\} ~|~ i=1,\dots,2k \} \\
      &\cup \{ \{ (2i,j-1), (2i,j)\} ~|~ i=1,\dots,k\text{ and }j=1,\dots,i \} \\
      &\cup \{ \{ (i+j-1,i), (i+j,i) \} ~|~ i=1,\dots,k\text{ and }j=1,\dots,i \}\text{.}
   \end{align*}
   This tree is visualized in Figure~\ref{fig:worstcaseexample}.
   We can easily see that a minimum-length Steiner tree has length $\text{SMT}_k = k^2 + 3k$, because the number of terminals is
   \begin{equation*}
      |T| = 2k + \sum_{i=1}^k 2i = 2k + k(k+1) = k^2 + 3k
   \end{equation*}
   and each additional terminal requires at least one edge of length 1 to connect it. This also shows that $A^{SMT} = (V,E)$ is indeed a minimum-length Steiner tree.

   Now we claim that the cost of an optimum solution is $\text{OPT} = 2k^2 + 3k$. This can be seen by  adding edges of the form $\{(i,i-1), (i,i)\}$ to  $A^{SMT}$ and removing $\{(2i,0), (2i,1)\}$ for $i=1,\dots,k$. The resulting tree for $k=3$ can be seen in Figure~\ref{fig:worstcaseexample_opt}. This preserves the length of $k^2 + 3k$. As every terminal $t$ with positive delay weight is connected through a shortest $r$-$t$-path, it leads to the optimum delay costs of
   \begin{equation*}
      \sum_{i=1}^k \frac{2i-1}{2i} 2i = \sum_{i=1}^k (2i-1) = k(k+1) - k = k^2\text{.}
   \end{equation*}

   Any algorithm that starts with $A^{SMT}$ cannot decrease the total cost by deleting edges and then reconnecting the resulting components individually to $r$.
   To see this, we may assume a counter example with a minimum number of components after edge deletion.
   Each component contains a terminal $t_i$ with $w(t_i)>0$, as unweighted  components cannot benefit from dis- and reconnection.
   Let $C$ be the disconnected component containing a terminal $t_i$ with $w(t_i)>0$ and $i$ minimum,
   and let $e$ be the edge whose deletion disconnected $C$ from the root component.
   Reconnecting $C$ to $r$ via a vertex on the x-axis or on the  vertical segment at x-position  $x=2i$ does not improve the delay cost nor the connection cost
   compared to not deleting $e$. So the best remaining way to reconnect $C$ is through an $r$-$t_i$ edge.
   Compared to keeping $e$, the delay cost decreases by at most $\frac{2i-1}{2i}(4i-2i) = 2i-1$, which equals the (additional) connection cost.
   Thus, keeping $e$ results in a solution with the same price. This contradicts the minimality of the counter example.

Finally,
\begin{equation*}
  \begin{array}{rl}
      \text{cost}(A^{SMT}) & = \text{length}(A^{SMT}) + \sum_{i=1}^{k} w(t_i) \text{dist}_{A^{SMT}}(r,t_i)\\
      &= k^2 + 3k + \sum_{i=1}^k \frac{2i-1}{2i} 4i = 3k^2 + 3k.
  \end{array}
\end{equation*}
Thus,  $\frac{\text{cost}(A^{SMT})}{\text{OPT}} = \frac{3k^2 + 3k}{2k^2 + 3k}$ and $ \lim_{k\to\infty} \frac{3k^2 + 3k}{2k^2 + 3k} = \frac{3}{2}$ as claimed.
\end{proof}

It is worth mentioning that the ratio $ \frac{3}{2} $ is not tight and can be improved to at least $ \frac{5+4\sqrt{2}}{7} \approx 1.5224 $.
Consider the horizontal segments of two edges on the $ x $-axis between two consecutive vertical segments in Figure \ref{fig:worstcaseexample}. By extending some of these segments to contain three instead of two edges  such that the average segment length converges to $ 1+\sqrt{2} $, we can slightly improve the ratio  to $\frac{5+4\sqrt{2}}{7}$ at the  cost of a significantly more involved technical analysis.

\section{Conclusion}
\label{sec:conclusion}

We significantly improve the approximation factor for the
\textsc{Uniform Cost-Distance Steiner Tree Problem}. For the
lower bound (\ref{eq:lower_bound}), the factor is best possible if the
minimum-length Steiner tree problem can be solved optimally.

This is achieved by an enhancement of the cut-off routine, where we do
not simply cut off by delay weight, but take the (cost)  structure of the sub-arborescences into account.  Furthermore, the root
component will be reconnected in a smarter way.
The analysis of our algorithm is tight if we start with minimum-length Steiner trees.

Our algorithm is very fast. After computing an approximate
minimum-length Steiner tree, the remaining cutting and re-assembling
takes linear time, which previously took a quadratic running time.

Any algorithm that follows the general strategy of cutting a Steiner tree into
pieces that are reconnected to the root  cannot achieve an approximation ratio better than $\frac{3}{2}$.

Based on our lower bound gap result and the lower bound on the approximation factor for any cut-and-reconnect algorithm, further attempts to improve the approximation ratio should improve the algorithm and the lower bound.

\bibliography{references}

\appendix


\section{Optimality Gap of Lower Bound in the Manhattan Plane}
\label{sec:gap-lb-manhattan}
In the Manhattan plane the lower bound gap is at least
$\sqrt{2}$ as the following family of instances shows. The
root is placed at $(0,0)$. Two terminals $t_1, t_2$ with  weight
$\frac{1}{\sqrt{2}}$ are placed at $(1,0)$ and $(0,1)$, respectively.
We place further unweighted uniformly spaced terminals terminals on the
lines $(0,0)--(1,1)$, $(1,0)--(1,1)$, and $(0,1)--(1,1)$ as indicated in Figure~\ref{fig:opt_lb_gap_manhattan}. In the family the spacing is decreased.
\begin{figure}
  \centering
  \begin{tikzpicture}[scale=0.7]
    \node[fsquare,red,label=$r$] (r) at (0,0) {};
    \node[fsquare,blue,label=left:{$t_1$}] (t1) at (5,0) {};
    \node[fsquare,blue,label=below:{$t_2$}] (t2) at (0,5) {};
    \node[fsquare,black,] (t3) at (5,5) {};
    \foreach \x in {1,...,19} {
      \node[fsquare,black,scale=0.5]  at (\x/4,\x/4) {};
      \node[fsquare,black,scale=0.5]  at (\x/4,5) {};
      \node[fsquare,black,scale=0.5]  at (5,\x/4) {};
      }
  \end{tikzpicture}
  \caption{An example from the family of instances attaining an optimum vs. lower bound gap of $\sqrt{2}$ in the Manhattan plane.}
    \label{fig:opt_lb_gap_manhattan}
\end{figure}

A minimum-length Steiner tree connects all terminals along these lines.
It has an asymptotic  connection cost  of $2+\sqrt{2}$ and an asymptotic
delay cost of $2(1+\sqrt{2})\frac{1}{\sqrt{2}}$, i.e.\ a total cost of $2(2+\sqrt{2})$.
for a sufficiently small spacing,  an optimum tree must use most of these edges, too.
Adding shortcuts anywhere between the diagonal path to the horizontal sub-path on the top or
vertical sub-path on the right increases  the connection by at least the savings in the delay cost.
Thus, it is also an optimum solution up to a vanishing error.

The lower bound is  $ (2+\sqrt{2}) +  2 \frac{1}{\sqrt{2}} = 2(1+\sqrt{2})$.
and the asymptotic  gap is $$\frac{2(2+\sqrt{2})}{2(1+\sqrt{2})} = \sqrt{2}.$$

\section{Proof of Lemma~\ref{lem:kh-21}}
\label{sec:proof-lem-kh21}

\begin{proof}(Lemma~\ref{lem:kh-21})
Note that Lemma 1 in \cite{Khazraei21}  states only the bound (\ref{eqn:kh_bound-3}).
The bounds (\ref{eqn:kh_bound-1}) and (\ref{eqn:kh_bound-2}) follow immediately from their proof, which  we will briefly sketch: We choose a terminal $ t\in T' $ randomly with probability $ p_t := \frac{w(t)}{W_{A'}} $ as the ``port'' vertex (only for the analysis). Then we obtain:
\begin{itemize}
\item The  expected cost of $(r,t)$ is  $\mathbb{E}(c(r,t)) = \sum _{t\in T'} p_t c(r,t) = \frac{1}{W_{A'}} D_{T'} $.

\item The (deterministic) connection cost within $ A' $ is  $ C_{A'} $.

\item The expected  effective  delay cost of $(r,t)$ is \[ \mathbb{E}(W_{A'} \cdot c(r,t)) = W_{A'} \cdot\sum _{t\in T'} p_t c(r,t) = D_{T'}.\]

\item The expected delay weight served by an edge $ (x,y)\in E(A') $ is
\[                        \frac{W_{A'_y}}{W_{A'}}\cdot (W_{A'} - W_{A'_y}) + \frac{W_{A'} - W_{A'_y}}{W_{A'}}\cdot W_{A'_y} = \frac{2 W_{A'_y} (W_{A'} - W_{A'_y})}{W_{A'}} \leq \frac{W_{A'}}{2},
  \]

  where $ A'_y $ is the sub-arborescence of $ A'-(x,y) $ containing $ y $. The formula reflects the expected component of  $(A'-(x,y))$
  in which  the port vertex is located.
                Summation over all edges in $ A' $ yields the following expected delay cost contribution of $E( A')$: \[  \sum _{e = (x,y)\in E(A')} \frac{2 W_{A'_y} (W_{A'} - W_{A'_y})}{W_{A'}} c(e) \leq \frac{W_{A'}}{2} C_{A'}. \]
\end{itemize}
The addition of these four terms gives the expected total cost of connecting $ A' $ to the root $r$, and
provides the bound in (\ref{eqn:kh_bound-1}). The deterministic best choice of the ``port'' vertex in Algorithm~\ref{alg:khazraei21} cannot be more expensive.
Now,  (\ref{eqn:kh_bound-2})  holds as (\ref{eqn:kh_bound-1}) is maximized for $W_{A'_y}= \frac{1}{2}W_{A'}$.
Finally,  (\ref{eqn:kh_bound-3})  follows as $W_{A'} \le 2\mu$ or $A'$ is a (heavy) singleton.
\end{proof}


\section{Proof of Lemma~\ref{lem:non-positive_functions}}
\label{sec:proof-lem-non-positive_functions}

\begin{proof}
        Note that the functions $ f $ and $ g $ differ only in the last factor. Actually, because of $ \frac{1}{a} - \frac{1}{\mu} < 0 $, $ \frac{1}{\frac{1}{a-b} - \frac{1}{\mu}} > 0 $ and $ \frac{\mu}{2} - \frac{2((a-b) - c) c}{a-b} \leq \frac{\mu}{2} $ we get $ f(a, b,c) \geq g(a,b,c) $ for all $ (a,b,c)\in X^\mu $, so it is sufficient to show $ f(a,b,c)\leq 0 $.

        Combining the first summands of $f$ in Definition~\ref{def:f-and-g}, and simplifying the third summand, we get
        \begin{align*}
                f (a,b,c)
                &= \frac{4ca - 4c^2 -\mu a}{2a} + \frac{(\mu - a)(a - b)}{a (\mu + b - a)} \left( \frac{\mu (a-b) - 4c(a-b-c)}{2(a-b)} \right) \\
                &= \frac{(\mu - a)(4ca - 4c^2 -\mu a) + b(4ca - 4c^2 -\mu a)}{2a(\mu +b-a)} \\
                &\quad + \frac{(\mu - a) \left( \mu (a-b) - 4c(a-b-c) \right)}{2a(\mu +b-a)} \\
                &= \frac{b(4ca - 4c^2 -\mu a) + b(\mu - a) (4c-\mu)}{2a(\mu +b-a)} \\
                &= \frac{b(4ca - 4c^2 -\mu a +4\mu c -\mu^2 -4ca +\mu a)}{2a(\mu +b-a)} \\
                &= - \frac{b(2c -\mu)^2}{2a(\mu +b-a)} \\
                &\leq 0\text{,}
        \end{align*}
        where $\mu + b - a > 0$ by $(a,b,c)\in X^{\mu}$.
\end{proof}

\section{Detailed Proof of Theorem~\ref{thm:running_time}}
\label{appendix:running-time}
\begin{proof}(Theorem~\ref{thm:running_time})
   We will proof two claims:
   \begin{enumerate}
      \item Checking whether a branch should be cut off at the traversed vertex can be done in $\mathcal{O}(1)$ time.
      \item Choosing the ports can be done in linear time.
   \end{enumerate}
   Then, we just observe that Step 2 traverses the initial tree once, which also needs linear time.

   Proof of Claim 1:
   We keep track of five values for each node $v$ and its corresponding sub-arborescence $ A_v := (A_0)_v $:
   $  W_v := W_{A_v}$, the weight inside $A_v$, $D_v := D_{A_v}$, the minimum possible delay cost of $A_v$, $C_v := C_{A_v}$, the connection cost of $A_v$,
    $S^1_v := \sum_{e = (p,q)\in E(A_v)} W_q (W_v - W_q) c(e)$ and $S^2_v := \sum_{e = (p,q)\in E(A_v)} W_q c(e)$.

   For leaves, we can compute these in constant time.
   For a node $v$ with only one child $x$ (because the other has been cut off), we can compute the values as follows:
       $W_v =  W_x$, $D_v = D_x$, $C_v = C_x + c(v,x)$,
   $S^1_v = S^1_x$, and $S^2_v = S^2_x + W_x c(v,x).$

   Whenever we consider a node $v$ with children $x$ and $y$, and $ A_x := (A_v)_x $ and $ A_y := (A_v)_y $, we can compute the values for $v$ like so:
      $W_v = W_x + W_y$,
      $D_v = D_x + D_y$,
      $C_v = C_x + c(v,x) + C_y + c(v,y)$,
   \begin{align*}
      S^1_v &= \sum_{e = (p,q)\in E(A_v)} W_q (W_v - W_q) c(e)\\
      &= \left( \sum_{e = (p,q)\in E(A_x)} W_q (W_v - W_q) c(e) \right) + W_x (W_v - W_x) c(v,x) \\
      &\quad + \left( \sum_{e = (p,q)\in E(A_y)} W_q (W_v - W_q) c(e) \right) + W_y (W_v - W_y) c(v,y) \\
      &= \left( \sum_{e = (p,q)\in E(A_x)} W_q (W_x - W_q) c(e) \right) + \left( \sum_{e = (p,q)\in E(A_x)} W_q W_y c(e) \right) \\
      &\quad + W_x (W_v - W_x) c(v,x) \\
      &\quad + \left( \sum_{e = (p,q)\in E(A_y)} W_q (W_y - W_q) c(e) \right) + \left( \sum_{e = (p,q)\in E(A_y)} W_q W_x c(e) \right) \\
      &\quad + W_y (W_v - W_y) c(v,y) \\
      &= S^1_x + W_y S^2_x + W_x (W_v - W_x) c(v,x) + S^1_y + W_x S^2_y + W_y (W_v - W_y) c(v,y),
   \end{align*}
   and
   \begin{equation*}
      S^2_v =  S^2_x + W_x c(v,x) + S^2_y + W_y c(v,y).
   \end{equation*}

   Proof of Claim 2:
   For each node $ v\in V(A') $, the cost when using $ v $ as the port is
   \begin{align*}
      \mathrm{cost}_v
      = c(r,v) + C_{A'}  + \sum _{t\in T_{A'}} w(t)\cdot (c(r,v) +  c(E(A'_{[v,t]}))).
   \end{align*}

   So for an edge $ e = (x,y)\in E(A') $ we have
   \begin{align*}
      \mathrm{cost}_x - \mathrm{cost}_y
      &= c(r,x) - c(r,y) + \sum _{t\in T_{A'}} w(t) (c(r,x) - c(r,y))  + \sum _{t\in T_{A'_y}} w(t) c(e)   \\
      &\quad - \sum _{t\in T_{A'}\backslash T_{A'_y}} w(t) c(e) \\
      &= (c(r,x) - c(r,y)) (1+W_{A'}) + c(e) W_{A'_y} - c(e) (W_{A'} - W_{A'_y}).
   \end{align*}

   This allows us to compute in constant time the  cost for choosing  $ y $ as the port from  the cost for choosing its parent $ x $ as the port. We take advantage of this property and first compute the  cost for using the root of $ A' $ as the port in $ \mathcal{O} (|E(A')| + |T_{A'}|) $. Then, we find the  find the best ``port'' vertex in a top-down traversal in the claimed linear time.
   \end{proof}

\section{Upper Bound on \texorpdfstring{$h(x,y)$}{h(x,y)}}
\label{appendix:bounding-h}

We prove that for $\beta \ge 1$, $x,y \ge 0, x+y > 0$
  \begin{align*}
    h(x,y)
    &:= \frac{\beta x + y + \sqrt{2}\sqrt{\beta x y}}{x + y}\\
    & = \beta + \frac{(1-\beta )y + \sqrt{2}\sqrt{\beta x y}}{x + y}\\
    & \le \beta + \frac{\beta}{\sqrt{\beta ^2 + 1} + \beta - 1}.
  \end{align*}
  For shorter notation, we set
        \[a  := \frac{\beta}{\sqrt{\beta ^2 + 1} + \beta - 1} > 0\]
        and get
        \begin{align*}
                \frac{\beta}{2a} + 1 - \beta - a
                &= \frac{1}{2} \sqrt{\beta ^2 + 1} - \frac{1}{2} (\beta - 1) - \frac{\beta}{\sqrt{\beta ^2 + 1} + \beta - 1} \\
                &= \frac{(\sqrt{\beta ^2 + 1} - (\beta - 1)) (\sqrt{\beta ^2 + 1} + (\beta - 1)) - 2\beta}{2 (\sqrt{\beta ^2 + 1} + \beta - 1)} \\
                &= 0.
        \end{align*}
        Therefore,
        \begin{align*}
                h(x,y)
                &= \beta + \frac{(1-\beta ) y + \sqrt{2}\sqrt{\beta xy}}{x + y} - \frac{\left( \frac{\beta}{2a} + 1-\beta - a\right) y}{x + y} \\
                &= \beta + \frac{\sqrt{2}\sqrt{\beta xy}}{x + y} - \frac{\left( \frac{\beta}{2a} - a\right) y}{x + y} \\
                &= \beta + a - \frac{ax}{x+y} + \frac{\sqrt{2}\sqrt{\beta xy}}{x + y} - \frac{\frac{\beta}{2a} y}{x + y} \\
                &= \beta + a - \frac{a}{x+y} \left( \sqrt{x} - \frac{\sqrt{\beta}}{\sqrt{2}a} \sqrt{y}\right) ^2.
        \end{align*}
        As $ a > 0 $, we obtain
        \begin{align*}
                h(x,y)
                \leq \beta + a
                = \beta + \frac{\beta}{\sqrt{\beta ^2 + 1} + \beta - 1}.
        \end{align*}\qed

\end{document}